\documentclass[preprint,showpacs,preprintnumbers,amsmath,amssymb]{revtex4}

\newcommand\rf[1]{(\ref{eq:#1})}
\newcommand\lab[1]{\label{eq:#1}}
\newcommand\nonu{\nonumber}
\newcommand\br{\begin{eqnarray}}
\newcommand\er{\end{eqnarray}}
\newcommand\be{\begin{equation}}
\newcommand\ee{\end{equation}}

\newcommand\foot[1]{\footnotemark\footnotetext{#1}}
\newcommand\lb{\lbrack}
\newcommand\rb{\rbrack}

\newcommand\llb{\left\lbrack}
\newcommand\rrb{\right\rbrack}

\newcommand\lcurl{\left\{}
\newcommand\rcurl{\right\}}
\renewcommand\({\left(}
\renewcommand\){\right)}
\newcommand\bv{\bigm\vert}               
\newcommand\bgv{\bigg\vert}              

\newcommand\bc{\begin{center}}
\newcommand\ec{\end{center}}

\newcommand\partder[2]{\frac{{\partial {#1}}}{{\partial {#2}}}}

\renewcommand\a{\alpha}

\renewcommand\d{\delta}
\newcommand\D{\Delta}

\newcommand\vareps{\varepsilon}
\newcommand\g{\gamma}
\newcommand\G{\Gamma}

\newcommand\h{\frac{1}{2}}
\renewcommand\k{\kappa}
\renewcommand\l{\lambda}
\renewcommand\L{\Lambda}
\newcommand\m{\mu}
\newcommand\n{\nu}

\newcommand\om{\omega}

\newcommand\vp{\varphi}
\renewcommand\P{\Phi}
\newcommand\pa{\partial}

\newcommand\pr{\prime}

\renewcommand\r{\rho}
\newcommand\s{\sigma}
\renewcommand\S{\Sigma}
\renewcommand\t{\tau}
\renewcommand\th{\theta}

\newcommand\wti{\widetilde}
\newcommand\twomat[4]{\left(\begin{array}{cc}  
{#1} & {#2} \\ {#3} & {#4} \end{array} \right)}

\newcommand\cA{{\mathcal A}}

\newcommand\cE{{\mathcal E}}
\newcommand\cF{{\mathcal F}}

\newcommand\cJ{{\mathcal J}}

\newcommand\cV{{\mathcal V}}

\newcommand{\ct}[1]{\cite{#1}}
\newcommand{\bib}[1]{\bibitem{#1}}
\newcommand\PRL[3]{\textsl{Phys. Rev. Lett.} \textbf{#1}, #3 (#2)}

\newcommand\PRD[3]{\textsl{Phys. Rev.} \textbf{D#1}, #3 (#2)}

\newcommand\PLB[3]{\textsl{Phys. Lett.} \textbf{#1B}, #3 (#2)}
\newcommand\CQG[3]{\textsl{Class. Quantum Grav.} \textbf{#1}, #3 (#2)}

\newcommand\AoP[3]{\textsl{Ann. of Phys.} \textbf{#1}, #3 (#2)}

\newcommand\IJMPA[3]{\textsl{Int. J. Mod. Phys.} \textbf{A#1}, #3 (#2)}

\newcommand\Xdot{\stackrel{.}{X}}
\newcommand\xdot{\stackrel{.}{x}}

\newcommand\rdot{\stackrel{.}{r}}

\newcommand\vpdot{\stackrel{.}{\varphi}}

\newcommand\tdot{\stackrel{.}{t}}
\newcommand\zdot{\stackrel{.}{z}}
\newcommand\etadot{\stackrel{.}{\eta}}

\begin{document}


\title{Spherically Symmetric and Rotating Wormholes\\
Produced by Lightlike Branes}

\author{E.I. Guendelman and A. Kaganovich}%
\email{guendel@bgumail.bgu.ac.il , alexk@bgumail.bgu.ac.il}
\affiliation{%
Department of Physics, Ben-Gurion University of the Negev \\
P.O.Box 653, IL-84105 ~Beer-Sheva, Israel
}%

\author{E. Nissimov and S. Pacheva}%
\email{nissimov@inrne.bas.bg , svetlana@inrne.bas.bg}
\affiliation{%
Institute for Nuclear Research and Nuclear Energy,
Bulgarian Academy of Sciences \\
Boul. Tsarigradsko Chausee 72, BG-1784 ~Sofia, Bulgaria
}%

\begin{abstract}
Lightlike $p$-branes (\textsl{LL-branes}) with dynamical (variable) tension
allow simple and elegant Polyakov-type and dual to it Nambu-Goto-like
world-volume action formulations. Here we first briefly describe the
dynamics of \textsl{LL-branes} as test objects in various physically
interesting gravitational backgrounds of black hole type, including rotating
ones. Next we show that \textsl{LL-branes} are the appropriate gravitational 
sources that provide proper matter energy momentum tensors in the Einstein 
equations of motion needed to generate traversable wormhole solutions, in particular,
self-consistent cylindrical rotating wormholes, with the \textsl{LL-branes}
occupying their throats. Here a major role is being played by the dynamical 
\textsl{LL-brane} tension which turns out to be negative but may be of arbitrary
small magnitude. As a particular solution we obtain traversable wormhole with 
Schwarzschild geometry generated by a \textsl{LL-brane} positioned at the
wormhole throat, which represents the correct consistent realization of the 
original Einstein-Rosen ``bridge'' manifold.
\end{abstract}

\pacs{11.25.-w, 04.70.-s, 04.50.+h}

\maketitle

\section{Introduction}

Lightlike branes (\textsl{LL-branes}) are very interesting dynamical systems
which play an important role in the description of various physically
important phenomena in general relativity, such as: (i) impulsive lightlike 
signals arising in cataclysmic astrophysical events \ct{barrabes-hogan};
(ii) the ``membrane paradigm'' \ct{membrane-paradigm} of black hole physics; 
(iii) the thin-wall approach to domain walls coupled to 
gravity \ct{Israel-66,Barrabes-Israel-Hooft}.

More recently, \textsl{LL-branes} became significant also in the context of
modern non-perturbative string theory, in particular, as the so called
$H$-branes describing quantum horizons (black hole and cosmological)
\ct{kogan-01}, as Penrose limits of baryonic $D(=$Dirichlet) branes
\ct{mateos-02}, \textsl{etc} (see also \ct{nonperturb-string}).

In the original papers \ct{Israel-66,Barrabes-Israel-Hooft} \textsl{LL-branes}
in the context of gravity and cosmology have been extensively studied from a 
phenomenological point of view, \textsl{i.e.}, by introducing them without specifying
the Lagrangian dynamics from which they may originate\foot{In a more recent paper 
\ct{barrabes-israel-05} brane actions in terms of their pertinent extrinsic geometry
have been proposed which generically describe non-lightlike branes, whereas the 
lightlike branes are treated as a limiting case.}. 
On the other hand, we have proposed in a series of recent papers 
\ct{LL-brane-main,inflation-all,our-WH} a new class of concise Lagrangian actions, 
providing a derivation from first principles of the \textsl{LL-brane} dynamics.

There are several characteristic features of \textsl{LL-branes} which drastically
distinguish them from ordinary Nambu-Goto branes: 

(i) They describe intrinsically lightlike modes, whereas Nambu-Goto branes describe
massive ones.

(ii) The tension of the \textsl{LL-brane} arises as an {\em additional
dynamical degree of freedom}, whereas Nambu-Goto brane tension is a given
{\em ad hoc} constant.

(iii) Consistency of \textsl{LL-brane} dynamics in a spherically or axially
symmetric gravitational background of codimension one requires the presence
of an event horizon which is automatically occupied by the \textsl{LL-brane}
(``horizon straddling'' according to the terminology of the first 
ref.\ct{Barrabes-Israel-Hooft}).

In the present paper we will explore the novel possibility of employing
\textsl{LL-branes} as natural self-consistent gravitational sources for 
wormhole space-times, in other words, generating wormhole solutions in
self-consistent bulk gravity-matter systems coupled to
\textsl{LL-branes} through dynamically derived world-volume \textsl{LL-brane}
stress energy tensors. For a review of wormhole space-times, 
see \ct{visser-book,WH-rev}.

The possibility of a ``wormhole space-time'' was first hinted at in the work
of Einstein and Rosen \ct{einstein-rosen}, where they considered matching at
the horizon of two identical copies of the exterior Schwarzschild space-time
region (subsequently called {\em Einstein-Rosen ``bridge''}).
The original Einstein-Rosen ``bridge'' manifold appears as a particular case of the
construction of spherically symmetric wormholes produced by
\textsl{LL-branes} as gravitational sources 
(refs.\ct{our-WH} and Section 4 below). 
The main lesson here is that consistency of Einstein
equations of motion yielding the original Einstein-Rosen ``bridge'' as well-defined
solution necessarily requires the presence of \textsl{LL-brane}
energy-momentum tensor as a source on the right hand side.
Thus, the introduction of \textsl{LL-brane} coupling to gravity 
brings the original Einstein-Rosen construction in ref.\ct{einstein-rosen} 
to a consistent completion (see the {\em Appendix} for details). 

Let us particularly emphasize that here and in what follows we consider the
Einstein-Rosen ``bridge'' in  its original formulation in ref.\ct{einstein-rosen}
as a four-dimensional space-time manifold consisting of two copies of the
exterior Schwarzschild space-time region matched along the horizon\foot{The 
nomenclature of 
``Einstein-Rosen bridge'' in several standard textbooks (\textsl{e.g.} \ct{MTW}) 
uses the Kruskal-Szekeres manifold. The latter notion of ``Einstein-Rosen bridge''
is not equivalent to the original construction in \ct{einstein-rosen}. Namely, 
the two regions in Kruskal-Szekeres space-time corresponding to the outer 
Schwarzschild space-time region ($r>2m$) and labeled $(I)$ and $(III)$ in \ct{MTW} 
are generally
{\em disconnected} and share only a two-sphere (the angular part) as a common border
($U=0, V=0$ in Kruskal-Szekeres coordinates), whereas in the original Einstein-Rosen
``bridge'' construction the boundary between the two identical copies of the
outer Schwarzschild space-time region ($r>2m$) is a three-dimensional hypersurface
($r=2m)$.}.

A more complicated example of a spherically symmetric wormhole with
Reissner-Nordstr{\"o}m geometry has also been presented in \ct{our-WH},
where two copies of the outer Reissner-Nordstr{\"o}m space-time region are
matched via \textsl{LL-brane} along what used to be the outer horizon of the
full Reissner-Nordstr{\"o}m manifold (see also Section 4 below). In this way we
obtain a wormhole solution which combines the features of the Einstein-Rosen
``bridge'' on the one hand (with wormhole throat at horizon), and the features of
Misner-Wheeler wormholes \ct{misner-wheeler}, \textsl{i.e.}, exhibiting the so called 
``charge without charge'' phenomenon\foot{Misner and Wheeler \ct{misner-wheeler}
realized that wormholes connecting two asymptotically flat space times provide 
the possibility of ``charge without charge'', \textsl{i.e.}, electromagnetically
non-trivial solutions where the lines of force of the electric field flow from one 
universe to the other without a source and giving the impression of being 
positively charged in one universe and negatively charged in the other universe.},
on the other hand.

In the present paper the results of refs.\ct{our-WH} will be extended to the case 
of rotating (and charged) wormholes. Namely, we will construct rotating cylindrically
symmetric wormhole solutions by matching two copies of the outer region of
rotating cylindrically symmetric (charged) black hole via rotating \textsl{LL-branes} 
sitting at the wormhole throat which in this case is the outer horizon of 
the corresponding rotating black hole.
Let us stress again that in doing so we will be solving Einstein equations
of motion systematically derived from a well-defined action principle,
\textsl{i.e.}, a Lagrangian action describing bulk gravity-matter system
coupled to a \textsl{LL-brane}, so that the energy-momentum tensor on the
r.h.s. of Einstein equations will contain as a crucial piece the
explicit world-volume stress-energy tensor of the \textsl{LL-brane} given by
the \textsl{LL-brane} world-volume action.

In Section 2 of the present paper we briefly review our construction of 
\textsl{LL-brane} world-volume actions for {\em arbitrary} world-volume dimensions. 

In Section 3 we discuss the properties of \textsl{LL-brane} dynamics as {\em
test} branes moving in generic spherically or axially symmetric gravitational 
backgrounds. In the present paper we concentrate on the special case of 
codimension one \textsl{LL-branes}. Here consistency of the \textsl{LL-brane} 
dynamics dictates that the bulk space-time must possess an event horizon which is 
automatically occupied by the \textsl{LL-brane} (``horizon straddling''). In
the case of rotating black hole backgrounds the test \textsl{LL-brane}
rotates along with the rotation of the horizon. Also, similarly to the
non-rotating case we find exponential ``inflation/deflation'' of the 
test \textsl{LL-brane}'s dynamical tension.

In Section 4 we consider self-consistent systems of bulk gravity and matter
interacting with \textsl{LL-branes}. We present the explicit construction of 
wormhole solutions to the Einstein equations with spherically symmetric or
rotating cylindrically symmetric geometry, generated through the pertinent 
\textsl{LL-brane} energy-momentum tensor. 

In Section 5 we briefly describe the traversability of the newly found
rotating cylindrical wormholes.

In the {\em Appendix} we first show that the Einstein-Rosen ``bridge'' solution 
in terms of the original coordinates introduced in \ct{einstein-rosen} 
{\em does not} satisfy the vacuum Einstein equations due to an ill-defined 
$\d$-function contribution at the throat appearing on the r.h.s. -- a would-be 
``thin shell'' matter energy-momentum tensor. Then we show how our present 
construction of wormhole solutions via \textsl{LL-branes} at their throats 
resolves the above problem and furnishes a satisfactory
completion of the original construction of Einstein-Rosen ``bridge''
\ct{einstein-rosen}. In other words, the fully consistent formulation of the 
original Einstein-Rosen ``bridge'' manifold as two identical copies of the exterior
Schwarzschild space-time region matched along the horizon must include a
gravity coupling to a \textsl{LL-brane}, which produces the proper surface
stress-energy tensor (derived from a well-defined world-volume Lagrangian)
necessary for the ``bridge'' metric to satify the pertinent Einstein
equations everywhere, including at the throat.

\section{Lightlike Branes: World-Volume Action Formulations}

In a series of previous papers \ct{LL-brane-main,inflation-all,our-WH} we proposed 
manifestly reparametrization invariant world-volume actions describing
intrinsically lightlike $p$-branes for any world-volume dimension $(p+1)$:
\be
S = - \int d^{p+1}\s \,\P (\vp)
\Bigl\lb \h \g^{ab} \pa_a X^{\m} \pa_b X^{\n} G_{\m\n}(X) - L\!\( F^2\)\Bigr\rb
\lab{LL-brane}
\ee
Here the following notions and notations are used:

\begin{itemize}
\item
Alternative non-Riemannian integration measure density $\P (\vp)$ (volume form) on
the $p$-brane world-volume manifold:
\be
\P (\vp) \equiv \frac{1}{(p+1)!} \vareps_{I_1\ldots I_{p+1}}
\vareps^{a_1\ldots a_{p+1}} \pa_{a_1} \vp^{I_1}\ldots \pa_{a_{p+1}} \vp^{I_{p+1}}
\lab{mod-measure-p}
\ee
instead of the usual $\sqrt{-\g}$. Here $\lcurl \vp^I \rcurl_{I=1}^{p+1}$ are
auxiliary world-volume scalar fields; $\g_{ab}$ ($a,b=0,1,{\ldots},p$)
denotes the intrinsic Riemannian metric on the world-volume, and
$\g = \det\Vert\g_{ab}\Vert$. Note that $\g_{ab}$ is {\em independent} of $\vp^I$.
The alternative non-Riemannian volume form \rf{mod-measure-p} has been first
introduced in the context of modified standard (non-lightlike) string and
$p$-brane models in refs.\ct{mod-measure}.
\item
$X^\m (\s)$ are the $p$-brane embedding coordinates in the bulk
$D$-dimensional space time with bulk Riemannian metric
$G_{\m\n}(X)$ with $\m,\n = 0,1,\ldots ,D-1$; 
$(\s)\equiv \(\s^0 \equiv \t,\s^i\)$ with $i=1,\ldots ,p$;
$\pa_a \equiv \partder{}{\s^a}$.
\item
$g_{ab}$ is the induced metric:
\be
g_{ab} \equiv \pa_a X^{\m} \pa_b X^{\n} G_{\m\n}(X) \; ,
\lab{ind-metric}
\ee
which becomes {\em singular} on-shell (manifestation of the lightlike nature, 
cf. Eq.\rf{gamma-eqs} below).
\item
Auxiliary $(p-1)$-rank antisymmetric tensor gauge field $A_{a_1\ldots a_{p-1}}$
on the world-volume with $p$-rank field-strength and its dual:
\be
F_{a_1 \ldots a_{p}} = p \pa_{[a_1} A_{a_2\ldots a_{p}]} \quad ,\quad
F^{\ast a} = \frac{1}{p!} \frac{\vareps^{a a_1\ldots a_p}}{\sqrt{-\g}}
F_{a_1 \ldots a_{p}}  \; .
\lab{p-rank}
\ee
Its Lagrangian $L\!\( F^2\)$ is {\em arbitrary} function of $F^2$ with the 
short-hand notation:
\be
F^2 \equiv F_{a_1 \ldots a_{p}} F_{b_1 \ldots b_{p}} 
\g^{a_1 b_1} \ldots \g^{a_p b_p} \; .
\lab{F2-id}
\ee
\end{itemize}

Let us note the simple identity:
\be
F_{a_1 \ldots a_{p-1}b}F^{\ast b} = 0 \; ,
\lab{F-id}
\ee
which will play a crucial role in the sequel.

\textbf{Remark 1.} For the special choice
$L\!\( F^2\)= \( F^2\)^{1/p}$ the action \rf{LL-brane} becomes 
manifestly invariant under {\em Weyl (conformal) symmetry}: 
$\g_{ab}\!\! \longrightarrow\!\! \g^{\pr}_{ab} = \rho\,\g_{ab}$,
$\vp^{I} \longrightarrow \vp^{\pr\, I} = \vp^{\pr\, I} (\vp)$ with Jacobian 
$\det \Bigl\Vert \frac{\pa\vp^{\pr\, I}}{\pa\vp^J} \Bigr\Vert = \rho$.
In what follows we will consider the generic Weyl {\em non}-invariant case.

Rewriting the action \rf{LL-brane} in the following equivalent form:
\be
S = - \int d^{p+1}\!\s \,\chi \sqrt{-\g}
\Bigl\lb \h \g^{ab} \pa_a X^{\m} \pa_b X^{\n} G_{\m\n}(X) - L\!\( F^2\)\Bigr\rb
\quad, \quad
\chi \equiv \frac{\P (\vp)}{\sqrt{-\g}}
\lab{LL-brane-chi}
\ee
with $\P (\vp)$ the same as in \rf{mod-measure-p},
we find that the composite field $\chi$ plays the role of a {\em dynamical
(variable) brane tension}. Let us note that the notion of dynamical brane 
tension has previously appeared in different contexts in refs.\ct{townsend-etal}.

Now let us consider the equations of motion corresponding to \rf{LL-brane} 
w.r.t. $\vp^I$:
\be
\pa_a \Bigl\lb \h \g^{cd} g_{cd} - L(F^2)\Bigr\rb = 0 \quad \longrightarrow \quad
\h \g^{cd} g_{cd} - L(F^2) = M  \; ,
\lab{phi-eqs}
\ee
where $M$ is an arbitrary integration constant. The equations of motion w.r.t.
$\g^{ab}$ read:
\be
\h g_{ab} - F^2 L^{\pr}(F^2) \llb\g_{ab} 
- \frac{F^{*}_a F^{*}_b}{F^{*\, 2}}\rrb = 0  \; ,
\lab{gamma-eqs}
\ee
where $F^{*\, a}$ is the dual field strength \rf{p-rank}. In deriving \rf{gamma-eqs}
we made an essential use of the identity \rf{F-id}.

Before proceeding, let us mention that both the auxiliary world-volume scalars 
$\vp^I$ entering the non-Riemannian integration measure density
\rf{mod-measure-p}, as well as the intrinsic world-volume metric $\g_{ab}$ are
{\em non-dynamical} degrees of freedom in the action \rf{LL-brane},
or equivalently, in \rf{LL-brane-chi}. Indeed, there are no (time-)derivatives
w.r.t. $\g_{ab}$, whereas the action \rf{LL-brane} (or \rf{LL-brane-chi}) is
{\em linear} w.r.t. the velocities $\stackrel{.}{\vp}^I$. Thus,
\rf{LL-brane} is a constrained dynamical system, \textsl{i.e.}, a system with 
gauge symmetries including the gauge symmetry under world-volume
reparametrizations, and 
both Eqs.\rf{phi-eqs}--\rf{gamma-eqs} are in fact
{\em non-dynamical constraint} equations (no second-order time derivatives
present). Their meaning as constraint
equations is best understood within the framework of the Hamiltonian formalism 
for the action \rf{LL-brane}. The latter can be developed
in strict analogy with the Hamiltonian formalism for a simpler class of
modified {\em non-lightlike} $p$-brane models based on the alternative non-Riemannian
integration measure density \rf{mod-measure-p}, which was previously proposed 
in \ct{m-string} (for details, we refer to Sections 2 and 3 of \ct{m-string}).
In particular, Eqs.\rf{gamma-eqs} can be viewed as $p$-brane 
analogues of the string Virasoro constraints.

There are two important consequences of Eqs.\rf{phi-eqs}--\rf{gamma-eqs}.
Taking the trace in \rf{gamma-eqs} and comparing with \rf{phi-eqs} 
implies the following crucial relation for the Lagrangian function $L\( F^2\)$: 
\be
L\!\( F^2\) - p F^2 L^\pr\!\( F^2\) + M = 0 \; ,
\lab{L-eq}
\ee
which determines $F^2$ \rf{F2-id} on-shell as certain function of the integration
constant $M$ \rf{phi-eqs}, \textsl{i.e.}
\be
F^2 = F^2 (M) = \mathrm{const} \; .
\lab{F2-const}
\ee

The second and most profound consequence of Eqs.\rf{gamma-eqs} is that the induced 
metric \rf{ind-metric} on the world-volume of the $p$-brane model \rf{LL-brane} 
is {\em singular} on-shell (as opposed to the induced metric in the case of 
ordinary Nambu-Goto branes):
\br
g_{ab}F^{*\, b}=0 \; ,
\lab{on-shell-singular}
\er
\textsl{i.e.}, the tangent vector to the world-volume $F^{*\, a}\pa_a X^\m$
is {\em lightlike} w.r.t. metric of the embedding space-time.
Thus, we arrive at the following important conclusion: every point on the 
surface of the $p$-brane \rf{LL-brane} moves with the speed of light
in a time-evolution along the vector-field $F^{\ast a}$ which justifies the
name {\em LL-brane} (Lightlike-brane) model for \rf{LL-brane}.

\textbf{Remark 2.} Let us stress the importance of introducing the alternative 
non-Riemannian integration measure density in the form \rf{mod-measure-p}.
If we would have started with world-volume \textsl{LL-brane} action in the
form \rf{LL-brane-chi} where the tension $\chi$ would be an {\em elementary} field
(instead of being function of the measure-density scalars), then variation
w.r.t. $\chi$ would produce second Eq.\rf{phi-eqs} with $M$ identically zero. This
in turn by virtue of the constraint \rf{L-eq} (with $M=0$) would require the
Lagrangian $L(F^2)$ to assume the special fractional function form from 
\textbf{Remark 1} above. This special case of Weyl-conformally invariant 
\textsl{LL-branes} has been discussed in our older papers 
(first two refs.\ct{LL-brane-main}).

Before proceeding let us point out that we can add \ct{LL-brane-main}
to the \textsl{LL-brane} action \rf{LL-brane} natural couplings to bulk Maxwell and 
Kalb-Ramond gauge fields. The latter do not affect Eqs.\rf{phi-eqs} and 
\rf{gamma-eqs}, so that the conclusions about on-shell constancy of $F^2$ 
\rf{F2-const} and the lightlike nature \rf{on-shell-singular} of the $p$-branes 
under consideration remain unchanged.

Further, the equations of motion w.r.t. world-volume gauge field 
$A_{a_1\ldots a_{p-1}}$ (with $\chi$ as defined in \rf{LL-brane-chi} and
accounting for the constraint \rf{F2-const}) read:
\be
\pa_{[a}\( F^{\ast}_{b]}\, \chi\) = 0  \; .
\lab{A-eqs}
\ee
They allow us to introduce the dual ``gauge'' potential $u$:
\be
F^{\ast}_{a} = \mathrm{const}\, \frac{1}{\chi} \pa_a u \; ,
\lab{u-def}
\ee
enabling us to rewrite Eq.\rf{gamma-eqs} (the lightlike constraint)
in terms of the dual potential $u$ in the form:
\be
\g_{ab} = \frac{1}{2a_0}\, g_{ab} - \frac{2}{\chi^2}\,\pa_a u \pa_b u \quad ,
\quad a_0 \equiv F^2 L^{\pr}\( F^2\)\bv_{F^2=F^2(M)} = \mathrm{const}
\lab{gamma-eqs-u}
\ee
($L^\pr(F^2)$ denotes derivative of $L(F^2)$ w.r.t. the argument $F^2$).
From \rf{u-def} and \rf{F2-const} we obtain the relation: 
\be
\chi^2 = -2 \g^{ab} \pa_a u \pa_b u \; ,
\lab{chi2-eq}
\ee
and the Bianchi identity $\nabla_a F^{\ast\, a}=0$ becomes:
\be
\pa_a \Bigl( \frac{1}{\chi}\sqrt{-\g} \g^{ab}\pa_b u\Bigr) = 0  \; .
\lab{Bianchi-id}
\ee

Finally, the $X^\m$ equations of motion produced by the \rf{LL-brane} read:
\be
\pa_a \(\chi \sqrt{-\g} \g^{ab} \pa_b X^\m\) + 
\chi \sqrt{-\g} \g^{ab} \pa_a X^\n \pa_b X^\l \G^\m_{\n\l}(X) = 0  \;
\lab{X-eqs}
\ee
where $\G^\m_{\n\l}=\h G^{\m\k}\(\pa_\n G_{\k\l}+\pa_\l G_{\k\n}-\pa_\k G_{\n\l}\)$
is the Christoffel connection for the external metric.

Now it is straightforward to prove that the system of equations 
\rf{chi2-eq}--\rf{X-eqs} for $\( X^\m,u,\chi\)$, which are equivalent to the 
equations of motion \rf{phi-eqs}--\rf{A-eqs},\rf{X-eqs} resulting from the 
original Polyakov-type \textsl{LL-brane} action \rf{LL-brane}, can be equivalently 
derived from the following {\em dual} Nambu-Goto-type world-volume action: 
\be
S_{\rm NG} = - \int d^{p+1}\s \, T 
\sqrt{- \det\Vert g_{ab} - \frac{1}{T^2}\pa_a u \pa_b u\Vert}  \; .
\lab{LL-action-NG}
\ee
Here $g_{ab}$ is the induced metric \rf{ind-metric};
$T$ is {\em dynamical} tension simply related to the dynamical tension 
$\chi$ from the Polyakov-type formulation \rf{LL-brane-chi} as
$T^2= \frac{\chi^2}{4a_0}$ with $a_0$ -- same constant as in \rf{gamma-eqs-u}.

In what follows we will consider the initial Polyakov-type form \rf{LL-brane} of the
\textsl{LL-brane} world-volume action. World-volume reparametrization invariance 
allows to introduce the standard synchronous gauge-fixing conditions:
\be
\g^{0i} = 0 \;\; (i=1,\ldots,p) \; ,\; \g^{00} = -1
\lab{gauge-fix}
\ee
Also, we will use a natural ansatz for the ``electric'' part of the 
auxiliary world-volume gauge field-strength:
\be
F^{\ast i}= 0 \;\; (i=1,{\ldots},p) \quad ,\quad \mathrm{i.e.} \;\;
F_{0 i_1 \ldots i_{p-1}} = 0 \; ,
\lab{F-ansatz}
\ee
meaning that we choose the lightlike direction in Eq.\rf{on-shell-singular} 
to coincide with the brane
proper-time direction on the world-volume ($F^{*\, a}\pa_a \sim \pa_\t$).
The Bianchi identity ($\nabla_a F^{\ast\, a}=0$) together with 
\rf{gauge-fix}--\rf{F-ansatz} and the definition for the dual field-strength
in \rf{p-rank} imply:
\be
\pa_0 \g^{(p)} = 0 \quad \mathrm{where}\;\; \g^{(p)} \equiv \det\Vert\g_{ij}\Vert \; .
\lab{gamma-p-0}
\ee
Then \textsl{LL-brane} equations of motion acquire the form 
(recall definition of $g_{ab}$ \rf{ind-metric}):
\be
g_{00}\equiv \Xdot^\m\!\! G_{\m\n}\!\! \Xdot^\n = 0 \quad ,\quad g_{0i} = 0 \quad ,\quad
g_{ij} - 2a_0\, \g_{ij} = 0
\lab{gamma-eqs-0}
\ee
(the latter are analogs of Virasoro constraints), where the $M$-dependent constant
$a_0$ (the same as in \rf{gamma-eqs-u}) must be strictly positive;
\be
\pa_i \chi = 0 \qquad (\mathrm{remnant ~of ~Eq.\rf{A-eqs}})\; ;
\lab{A-eqs-0}
\ee

\vspace{-0.6cm}
\br
-\sqrt{\g^{(p)}} \pa_0 \(\chi \pa_0 X^\m\) +
\pa_i\(\chi\sqrt{\g^{(p)}} \g^{ij} \pa_j X^\m\)
\nonu \\
+ \chi\sqrt{\g^{(p)}} \(-\pa_0 X^\n \pa_0 X^\l + \g^{kl} \pa_k X^\n \pa_l X^\l\)
\G^\m_{\n\l} = 0 \; .
\lab{X-eqs-0}
\er

\section{Lightlike Test-Branes in Spherically and Axially Symmetric 
Gravitational Backgrounds}

First, let us consider codimension one \textsl{LL-brane} moving in a general
spherically symmetric background:
\be
ds^2 = - A(t,r)(dt)^2 + B (t,r) (dr)^2 + C(t,r) h_{ij}(\vec{\th}) d\th^i d\th^j \; ,
\lab{spherical-metric}
\ee
\textsl{i.e.}, $D=(p+1)+1$, with the simplest non-trivial ansatz for the 
\textsl{LL-brane} embedding coordinates $X^\m (\s)$: 
\be
t = \t \equiv \s^0 \;\; , \;\; r= r(\t) \;\; , \;\; \th^i = \s^i \;
(i=1,{\ldots},p) \; .
\lab{X-embed}
\ee
The \textsl{LL-brane} equations of motion \rf{gamma-eqs-u}--\rf{X-eqs}, taking into 
account \rf{gauge-fix}--\rf{F-ansatz}, acquire the form: 
\br
-A + B \rdot^2 = 0 \;\; ,\; \mathrm{i.e.}\;\; \rdot = \pm \sqrt{\frac{A}{B}}
\quad ,\quad
\pa_t C + \rdot \pa_r C = 0
\lab{r-const} \\
\pa_\t \chi + \chi \llb \pa_t \ln \sqrt{AB} 
\pm \frac{1}{\sqrt{AB}} \Bigl(\pa_r A + p\, a_0 \pa_r \ln C\Bigr)\rrb_{r=r(\t)} = 0
\; ,
\lab{X0-eq-1}
\er
where $a_0$ is the same constant appearing in \rf{gamma-eqs-u}.
In particular, we are interested in static spherically symmetric metrics in 
standard coordinates:
\be
ds^2 = - A(r)(dt)^2 + A^{-1}(r) (dr)^2 + r^2 h_{ij}(\vec{\th}) d\th^i d\th^j
\lab{standard-spherical}
\ee
for which Eqs.\rf{r-const} yield:
\be
\rdot = 0 \;\; ,\;\; \mathrm{i.e.}\;\; r(\t) = r_0 = \mathrm{const} \quad, \quad
A(r_0) = 0 \; .
\lab{horizon-standard}
\ee
Eq.\rf{horizon-standard} tells us that consistency of \textsl{LL-brane} dynamics in 
a spherically symmetric gravitational background of codimension one requires the 
latter to possess a horizon (at some $r = r_0$), which is automatically occupied 
by the \textsl{LL-brane} (``horizon straddling''). Further, Eq.\rf{X0-eq-1}
implies for the dynamical tension:
\be
\chi (\t) = \chi_0 
\exp\lcurl\mp \t \(\pa_r A\bv_{r=r_0} + \frac{2 p\, a_0}{r_0}\)\rcurl
\quad ,\quad \chi_0 = \mathrm{const} \; .
\lab{chi-eq-standard-sol}
\ee
Thus, we find a time-asymmetric solution for the dynamical
brane tension which (upon appropriate choice of the signs $(\mp)$ depending on the 
sign of the constant factor in the exponent on the r.h.s. of \rf{chi-eq-standard-sol})
{\em exponentially ``inflates'' or ``deflates''} for large times (for details 
we refer to \ct{inflation-all}). This phenomenon is an analog of the 
``mass inflation'' effect around black hole horizons \ct{poisson-israel}.

Next, let us consider $D\! =\! 4$-dimensional Kerr-Newman background metric in the 
standard Boyer-Lindquist coordinates (see \textsl{e.g.} \ct{textbooks-kerr}):
\be
ds^2 = -A (dt)^2 - 2 E dt\,d\vp + \frac{\S}{\D} (dr)^2 + \S (d\th)^2 +
D \sin^2 \th (d\vp)^2  \; ,
\lab{kerr-metric}
\ee
\be
A\equiv \frac{\D - a^2 \sin^2 \th}{\S} \;\; ,\;\;
E\equiv \frac{a \sin^2 \th\,\(r^2 + a^2 - \D\)}{\S} \;\; ,\;\;
D\equiv \frac{\( r^2 + a^2\)^2 - \D a^2 \sin^2 \th}{\S} \; ,
\lab{kerr-coeff}
\ee
where $\S \equiv r^2 + a^2\cos^2 \th\; ,\; \D \equiv r^2 + a^2 + e^2 - 2 mr$.
Let us recall that the Kerr-Newman metric \rf{kerr-metric}--\rf{kerr-coeff}
reduces to the Reissner-Nordstr{\"o}m metric in the limiting case $a=0$.

For the \textsl{LL-brane} embedding we will use the following ansatz:
\be
X^0 \equiv t = \t \;\; ,\;\; r=r(\t) \;\; ,\;\; \th = \s^1 \;\; ,\;\;
\vp = \s^2 + {\wti \vp}(\t) \; .
\lab{kerr-ansatz}
\ee
In this case the \textsl{LL-brane} equations of motion
\rf{gamma-p-0}--\rf{gamma-eqs-0} acquire the form:
\br
-A + \frac{\S}{\D} \rdot^2 + D \sin^2 \th\,\vpdot^2 - 2 E \vpdot = 0
\nonu \\
-E + D \sin^2 \th\,\vpdot = 0 \quad ,\quad  
\frac{d}{d\t}\( D\S \sin^2 \th\) = 0 \; .
\lab{gamma-eqs-kerr}
\er
Inserting the ansatz \rf{kerr-ansatz} into \rf{gamma-eqs-kerr} the last 
Eq.\rf{gamma-eqs-kerr} implies:
\be
r(\t) = r_0 = \mathrm{const} \; , 
\lab{kerr-horizon}
\ee
whereas the second Eq.\rf{gamma-eqs-kerr} yields:
\be
\D (r_0) = 0 \quad ,\quad \om \equiv \vpdot = \frac{a}{r_0^2 + a^2}
\lab{kerr-dragging}
\ee
Eqs.\rf{kerr-horizon}--\rf{kerr-dragging} indicate that:\\
$\phantom{aa}$(i) the \textsl{LL-brane} 
automatically locates itself on the Kerr-Newman horizon $r=r_0$ -- horizon 
``straddling'' according to the terminology of the first 
ref.\ct{Barrabes-Israel-Hooft};\\
$\phantom{aa}$(ii) the \textsl{LL-brane}
rotates along with the same angular velocity $\om$ as the Kerr-Newman horizon.

The first Eq.\rf{gamma-eqs-kerr} implies that $\rdot$ vanishes
on-shell as:
\be
\rdot\, \simeq\, \pm \frac{\D (r)}{r_0^2 + a^2}\bv_{r \to r_0} \; .
\lab{r-dot}
\ee
We will also need the explicit form of the last Eq.\rf{gamma-eqs-0} (using notations 
\rf{kerr-coeff}):
\be
\g_{ij} = \frac{1}{2a_0} \twomat{\S}{0}{0}{D \sin^2 \th}\bgv_{r=r_0 ,\th=\s^1} \; .
\lab{g-ij}
\ee
Among the $X^\m$-equations of motion \rf{X-eqs-0} only the $X^0$-equation
yields additional information. Because of the embedding $X^0 = \t$ it acquires 
the form of a time-evolution equation for the dynamical brane tension $\chi$:
\be
\pa_\t \chi + \chi\,\llb \pa_\t X^\n \pa_\t X^\l 
- \g^{ij} \pa_i X^\n \pa_j X^\l \rrb \G^{0}_{\n\l} = 0 \; ,
\lab{X0-eq}
\ee
which, after taking into account \rf{kerr-ansatz},
\rf{kerr-horizon}--\rf{kerr-dragging}
and the explicit expressions for the Kerr-Newman Christoffel connection 
coefficients (first ref.\ct{textbooks-kerr}), reduces to:
\be
\pa_\t \chi + \chi\, 2\rdot\Bigl\lb \G^0_{0r} + 
\frac{a}{r_0^2 + a^2}\G^0_{r\vp}\Bigr\rb_{r=r_0} = 0 \; .
\lab{X0-eq-1-kerr}
\ee
Singularity on the horizon of the Christoffel coefficients $\(\sim \D^{-1}\)$
appearing in \rf{X0-eq-1-kerr} is cancelled by $\D$ in $\rdot$ \rf{r-dot} so that 
finally we obtain:
\be
\pa_\t \chi \pm \chi \frac{2(r_0 - m)}{r_0^2 + a^2} = 0 \;\; ,\; \mathrm{i.e.}\;\;
\chi = \chi_0 \exp\Bigl\{\mp 2 \frac{(r_0 - m)}{r_0^2 + a^2}\,\t\Bigr\}
\lab{chi-eq-kerr}
\ee
Thus, we find ``mass inflation/deflation'' effect (according to the
terminology of \ct{poisson-israel}) on the Kerr-Newman horizon via the exponential time
dependence of the dynamical \textsl{LL-brane} tension similar to the 
``mass inflation/deflation'' effect with \textsl{LL-branes} in spherically symmetric
gravitational backgrounds (Eq.\rf{chi-eq-standard-sol}).

Now let us consider rotating cylindrical black hole background in $D=4$
\ct{lemos-etal}:
\be
ds^2 = -A (dt)^2 - 2 E dt\,d\vp + \frac{(dr)^2}{\D} + D (d\vp)^2 + \a^2 r^2 (dz)^2 
\lab{rot-cylindr-metric}
\ee
where:
\be
A \equiv -\om^2 r^2 + \g^2 \D \;\; ,\;\; E \equiv \g\om r^2 - \frac{\g\om}{\a^2}\D 
\;\;, \;\; D \equiv \g^2 r^2 - \frac{\om^2}{\a^4}\D \;\;, \;\;
\D \equiv \a^2 r^2 - \frac{b}{\a r} +\frac{c^2}{\a^2 r^2} \; .
\lab{rot-cylindr-coeff}
\ee
The physical meaning of the parameters involved is as follows:
$\a^2 = - \frac{1}{3}\L$, \textsl{i.e.}, $\L$ must be negative cosmological
constant; $b=4m$ with $m$ being the mass per unit length along the $z$-axis;
$c^2 = 4 \l^2$, where $\l$ indicates the linear charge density along the $z$-axis.

The metric \rf{rot-cylindr-coeff} possesses in general two horizons at
$r=r_{(\pm)}$ where $\D\bigl( r_{(\pm)}\bigr) = 0$.

Let us note the useful identity which will play an important role in the
sequel:
\be
AD+E^2 = r^2 \Bigl(\g^2 - \frac{\om^2}{\a^2}\Bigr)^2\,\D
\lab{ADE2-id}
\ee

For the \textsl{LL-brane} embedding we will use an ansatz similar to the
Kerr-Newman case \rf{kerr-ansatz}:
\be
X^0 \equiv t = \t \;\; ,\;\; r=r(\t) \;\; ,\;\; z = \s^1 \;\; ,\;\;
\vp = \s^2 + {\wti \vp}(\t) \; .
\lab{cylindr-ansatz}
\ee
Then the lightlike and Virasoro-like constraint equations of the \textsl{LL-brane}
dynamics \rf{gamma-eqs-0} in the background 
\rf{rot-cylindr-metric}--\rf{rot-cylindr-coeff} (the analogs of Eqs.\rf{gamma-eqs-kerr}
in the Kerr-Newman case):
\be
-A + \frac{\rdot^2}{\D} + D\,\vpdot^2 - 2 E \vpdot = 0  \quad ,\quad 
-E + D \,\vpdot = 0 \quad ,\quad \frac{d}{d\t}\( D\,\a^2r^2\) = 0
\lab{gamma-eqs-cylindr}
\ee
imply:
\be
r(\t) = r_0 = \mathrm{const} \;\;, \;\; \D (r_0) = 0  \;\;, \;\; 
\vpdot = \frac{\om}{\g} \; .
\lab{cylindr-horizon-dragging}
\ee
Thus, similarly to the Kerr-Newman case:\\
$\phantom{aa}$(i) the \textsl{LL-brane} 
automatically locates itself on one of the cylindrical black hole horizons at
$r = r_0 =r_{(\pm)}$ (horizon ``straddling'');\\
$\phantom{aa}$(ii) the \textsl{LL-brane} rotates with angular velocity $\om/\g$ 
along with the rotation of the cylindrical black hole horizon;\\
$\phantom{aa}$(iii) $\rdot$ vanishes on-shell as:
\be
\rdot\, \simeq \pm \bgv \frac{1}{\g}\Bigl(\g^2 
- \frac{\om^2}{\a^2}\Bigr)\bgv\, \D(r)\bv_{r \to r_{(\pm)}}
\lab{r-dot-cylindr}
\ee
(the overall signs $\pm$ on the r.h.s. of \rf{r-dot-cylindr} are {\em not} 
correlated with the indices $(\pm)$ labelling the outer/inner horizon).

Again in complete analogy with the Kerr-Newman case
(Eqs.\rf{X0-eq}--\rf{X0-eq-1-kerr}) the $X^0$-equation of motion \rf{X-eqs-0}
reduces to the following time-evolution equation for the pertinent dynamical brane
tension:
\be
\pa_\t \chi + \chi\, 2\rdot\Bigl\lb \G^0_{0r} + 
\frac{E}{D}\G^0_{r\vp}\Bigr\rb_{r=r_{(\pm)}} = 0 \; ,
\lab{X0-eq-1-cylindr}
\ee
where the Christoffel coefficients read:
\be
\G^0_{0r} = \frac{D \pa_r A + E \pa_r E}{2(AD+E^2)} \quad ,\quad
\G^0_{r\vp} = \frac{D \pa_r E - E \pa_r D}{2(AD+E^2)}
\lab{christoffel-cylindr}
\ee
with the functions $A,D,E$ as in \rf{rot-cylindr-metric}--\rf{rot-cylindr-coeff}.
Taking into account \rf{cylindr-horizon-dragging}--\rf{r-dot-cylindr} and the 
identity \rf{ADE2-id} we obtain from \rf{X0-eq-1-cylindr}--\rf{christoffel-cylindr}
exponential ``inflation/deflation'' of the \textsl{LL-brane} tension in
rotating cylindrical black hole background:
\be
\chi (\t) = 
\chi_0 \exp\lcurl\mp\t\,\bgv\frac{1}{\g}\Bigl(\g^2-\frac{\om^2}{\a^2}\Bigr)\bgv\, 
\pa_r \D(r)\bv_{r=r_{(\pm)}}\rcurl
\lab{chi-eq-cylindr}
\ee
(here again there is no correllation between the overall signs $\mp$ in the
exponent with the indices $(\pm)$ labelling the outer/inner horizon).
 
\section{Self-Consistent Wormhole Solutions via Lightlike Branes}

Let us now consider a self-consistent bulk Einstein-Maxwell system (with a
cosmological constant) free of electrically charged matter, coupled to a 
codimension one \textsl{LL-brane}:
\be
S = \int\!\! d^D x\,\sqrt{-G}\,\llb \frac{R(G)}{16\pi} - \frac{\L}{8\pi} 
- \frac{1}{4} \cF_{\m\n}\cF^{\m\n}\rrb 
+ S_{\mathrm{LL}} \; .
\lab{E-M-LL}
\ee
Here $\cF_{\m\n} = \pa_\m \cA_\n - \pa_\n \cA_\m$ and $S_{\mathrm{LL}}$ is
the same \textsl{LL-brane} world-volume action as in \rf{LL-brane-chi}.
Thus, the \textsl{LL-brane} will serve as a gravitational source 
through its energy-momentum tensor (see Eq.\rf{T-brane} below).
The pertinent Einstein-Maxwell equations of motion read:
\be
R_{\m\n} - \h G_{\m\n} R + \L G_{\m\n} =
8\pi \( T^{(EM)}_{\m\n} + T^{(brane)}_{\m\n}\) \quad, \quad
\pa_\n \(\sqrt{-G}G^{\m\k}G^{\n\l} \cF_{\k\l}\) = 0 \; ,
\lab{Einstein-Maxwell-eqs}
\ee
where $T^{(EM)}_{\m\n} = \cF_{\m\k}\cF_{\n\l} G^{\k\l} - G_{\m\n}\frac{1}{4}
\cF_{\r\k}\cF_{\s\l} G^{\r\s}G^{\k\l}$, 
and the \textsl{LL-brane} energy-momentum tensor is straightforwardly derived
from \rf{LL-brane-chi}:
\be
T^{(brane)}_{\m\n} = - G_{\m\k}G_{\n\l}
\int\!\! d^{p+1} \s\, \frac{\d^{(D)}\bigl(x-X(\s)\bigr)}{\sqrt{-G}}\,
\chi\,\sqrt{-\g} \g^{ab}\pa_a X^\k \pa_b X^\l  \; ,
\lab{T-brane}
\ee
The equations of motion of the \textsl{LL-brane} have already been given in 
\rf{gamma-eqs-0}--\rf{X-eqs-0}.

Using \rf{T-brane} we will now construct traversable {\em wormhole} solutions to the 
Einstein equations \rf{Einstein-Maxwell-eqs} which will combine the features
of the Einstein-Rosen ``bridge'' (wormhole throat at horizon) and the feature
``charge without charge'' of Misner-Wheeler wormholes \ct{misner-wheeler}.
In doing this we will follow the standard procedure described in
\ct{visser-book}, but with the significant difference that in our case
we will solve Einstein equations following from a self-consistent bulk 
gravity-matter system coupled to a \textsl{LL-brane}.
In other words, the \textsl{LL-brane} will serve as a gravitational source
of the wormhole by locating itself on its throat as a result
of its consistent world-volume dynamics (Eq.\rf{horizon-standard} above). 

First we will consider the case with {\em spherical symmetry}.
To this end let us take a spherically symmetric solution of \rf{Einstein-Maxwell-eqs}
of the form \rf{standard-spherical} in the absence of the \textsl{LL-brane}  
(\textsl{i.e.}, without $T^{(brane)}_{\m\n}$ on the r.h.s.), which possesses an 
(outer) event horizon at some $r=r_0$ (\textsl{i.e.}, $A(r_0)=0$ and
$A(r)>0$ for $r>r_0$). At this point we 
introduce the following modification of the metric \rf{standard-spherical}:
\be
ds^2 = - {\wti A}(\eta)(dt)^2 + {\wti A}^{-1}(\eta) (d\eta)^2 + 
(r_0 + |\eta|)^2 h_{ij}(\vec{\th}) d\th^i d\th^j \quad ,\quad
{\wti A}(\eta) \equiv A(r_0 + |\eta|) \;\;, 
\lab{spherical-WH}
\ee
where $-\infty < \eta < \infty$.
From now on the bulk space-time indices $\m,\n$ will refer to $(t,\eta,\th^i)$
instead of $(t,r,\th^i)$.
The new metric \rf{spherical-WH} represents two identical copies of the
exterior region ($r > r_0$) of the spherically symmetric space-time with
metric \rf{standard-spherical}, which are sewed together along the horizon
$r=r_0$. We will show that the new metric \rf{spherical-WH} is a solution of the full
Einstein equations \rf{Einstein-Maxwell-eqs}, {\em including} 
$T^{(brane)}_{\m\n}$ on the r.h.s.. Here the newly introduced coordinate $\eta$
will play the role of a radial-like coordinate normal w.r.t. the
\textsl{LL-brane} located on the horizon, which interpolates between two copies of
the exterior region of \rf{standard-spherical} (the two copies transform into
each other under the ``parity'' transformation $\eta \to - \eta$).

Inserting in \rf{T-brane} the expressions for $X^\m (\s)$ from
\rf{X-embed} and \rf{horizon-standard} and taking into account \rf{gamma-eqs-u},
\rf{gauge-fix}--\rf{F-ansatz} we get:
\be
T_{(brane)}^{\m\n} = S^{\m\n}\,\d (\eta)
\lab{T-S-0}
\ee
with surface energy-momentum tensor:
\be
S^{\m\n} \equiv - \frac{\chi}{(2a_0)^{p/2}}\,
\llb - \pa_\t X^\m \pa_\t X^\n + \g^{ij} \pa_i X^\m \pa_j X^\n 
\rrb_{t=\t,\,\eta=0,\,\th^i =\s^i} \quad , \;\; \pa_i \equiv \partder{}{\s^i} \; ,
\lab{T-S-brane}
\ee
where again $a_0$ is the integration constant parameter appearing in the 
\textsl{LL-brane} dynamics (cf. Eq.\rf{gamma-eqs-u}). Let us also note that 
unlike the case of test \textsl{LL-brane} moving in a spherically symmetric
background (Eqs.\rf{X0-eq-1} and \rf{chi-eq-standard-sol}), the dynamical brane 
tension $\chi$ in Eq.\rf{T-S-brane} turns out to be {\em constant}. 
This is due to the fact 
that in the present context we have a discontinuity in the Christoffel connection 
coefficients across the \textsl{LL-brane} sitting on the horizon ($\eta = 0$). 
The problem in treating the geodesic \textsl{LL-brane} equations of motion
\rf{X-eqs}, in particular -- Eq.\rf{X0-eq-1}, can be resolved following the approach 
in ref.\ct{Israel-66} (see also the regularization approach in ref.\ct{BGG},
Appendix A) by taking the mean value of the pertinent non-zero Christoffel 
coefficients across the discontinuity at $\eta = 0$. From the explicit form of
Eq.\rf{X0-eq-1} it is straightforward to conclude that the above mentioned
mean values around $\eta = 0$ vanish since now $\pa_r$ is replaced by $\pa/\pa \eta$,
whereas the metric coefficients depend explicitly on $|\eta|$. Therefore, in the 
present case Eq.\rf{X0-eq-1} is reduced to $\pa_\t \chi = 0$.

Let us now separate in \rf{Einstein-Maxwell-eqs}
explicitly the terms contributing to $\delta$-function singularities (these
are the terms containing second derivatives w.r.t. $\eta$, bearing in
mind that the metric coefficients in \rf{spherical-WH} depend on $|\eta|$):
\br
R_{\m\n} \equiv \pa_\eta \G^{\eta}_{\m\n} - \pa_\m \pa_\n \ln \sqrt{-G}
+ \mathrm{non-singular ~terms}
\nonu \\
= 8\pi \( S_{\m\n} - \frac{1}{p} G_{\m\n} S^{\l}_{\l}\) \d (\eta) 
+ \mathrm{non-singular ~terms} \; .
\lab{E-M-eqs}
\er
The only non-zero contribution to the $\delta$-function singularities on
both sides of Eq.\rf{E-M-eqs} arises for $(\m\n)=(\eta \eta)$. In order to
avoid coordinate singularity on the horizon it is more convenient to
consider the mixed component version of the latter 
(with one contravariant and one covariant index) : 
\be
R^{\eta}_{\eta} = 8\pi \( S^{\eta}_{\eta} - \frac{1}{p} S^{\l}_{\l}\) \d (\eta) 
+ \mathrm{non-singular ~terms}
\lab{E-M-eqs-eta}
\ee
(in the special case of Schwarzschild geometry we can use the
Eddington-Finkelstein coordinate system which is free of singularities on
the horizon; see the {\em Appendix}).
Evaluating the l.h.s. of \rf{E-M-eqs-eta} through the formula (recall $D=p+2$): 
\be
R^r_r = - \h \frac{1}{r^{D-2}}\pa_r \( r^{D-2} \pa_r A\)
\lab{R-rr}
\ee
valid for any spherically symmetric metric of the form \rf{standard-spherical}
and recalling $r=r_0 + |\eta|$,
we obtain the following matching condition for the coefficients in front of
the $\delta$-functions on both sides of \rf{E-M-eqs-eta} (analog of Israel 
junction conditions \ct{Israel-66,Barrabes-Israel-Hooft}):
\be
\pa_\eta {\wti A}\bv_{\eta \to +0} - \pa_\eta {\wti A}\bv_{\eta \to -0} =
- \frac{16 \pi\,\chi}{(2a_0)^{p/2-1}} \; , 
\lab{israel-junction}
\ee
or, equivalently:
\be
\pa_r A \bv_{r=r_0} = - \frac{8 \pi\,\chi}{(2a_0)^{p/2-1}} \; ,
\lab{israel-junction-1}
\ee
where we have used the explicit expression for the trace of the
\textsl{LL-brane} energy-momentum tensor \rf{T-S-brane}:
\be
S^\l_\l = - \frac{p}{(2a_0)^{p/2-1}}\,\chi \; .
\lab{T-S-brane-trace}
\ee

Eq.\rf{israel-junction-1} yields a relation between the parameters of the
spherically symmetric outer regions of ``vacuum'' solution \rf{standard-spherical}
of Einstein Eqs.\rf{Einstein-Maxwell-eqs} and the dynamical tension of the 
\textsl{LL-brane} sitting at the (outer) horizon.

As an explicit example let us take \rf{standard-spherical} to be the
standard $D=4$ Reissner-Nordstr{\"o}m metric, \textsl{i.e.},
$A(r)= 1 - \frac{2m}{r} + \frac{e^2}{r^2}$. Then Eq.\rf{israel-junction}
yields the following relation between the Reissner-Nordstr{\"o}m
parameters and the dynamical \textsl{LL}-brane tension:
\be
4\pi\chi\,r_0^2 + r_0 - m = 0 \quad, \;\;\mathrm{where} \;\;
r_0 = m +\sqrt{m^2 -e^2} \; .
\lab{parameter-matching}
\ee
Eq.\rf{parameter-matching} indicates that the dynamical brane tension must
be {\em negative}. Eq.\rf{parameter-matching} reduces to a cubic equation
for the Reissner-Nordstr{\"o}m mass $m$ as function of $|\chi|$:
\be
\bigl( 16\pi\,|\chi|\, m - 1\bigr) \( m^2 - e^2\) + 16\pi^2 \chi^2 e^4 = 0 \; .
\lab{M-RN}
\ee
In the special case of Schwarzschild wormhole ($e^2 = 0$) the Schwarzschild mass
becomes:
\be
m = \frac{1}{16\pi\,|\chi|} \;\; .
\lab{M-Schw}
\ee
The particular case of Schwarzschild wormhole (with 
${\wti A}(\eta) = 1 - \frac{2m}{2m + |\eta|}$) constructed above is the proper
consistent realization of the Einstein-Rosen ``bridge'' \ct{einstein-rosen}.
We refer to the {\em Appendix}, where it is explained how the present
formalism involving a \textsl{LL-brane} as wormhole source positioned at
the wormhole throat resolves certain inconsistency in the original treatment of
the Einstein-Rosen ``bridge''.

Let us observe that for large values of the \textsl{LL-brane} tension $|\chi|$, 
the Reissner-Nordstr{\"o}m (Schwarzschild) mass $m$ is very small. In particular, 
$m << M_{Pl}$ for $|\chi| > M_{Pl}^3$ 
($M_{Pl}$ being the Planck mass). On the other hand, for small values of the 
\textsl{LL-brane} tension $|\chi|$ Eq.\rf{parameter-matching} implies that the 
Reissner-Nordstr{\"o}m geometry of the wormhole must be near extremal 
($m^2 \simeq e^2$).

Now we will apply the above formalism to construct a rotating cylindrically
symmetric wormhole in $D=4$. Namely, we introduce the following modification 
of the cylindrically symmetric rotating black hole metric 
\rf{rot-cylindr-metric}--\rf{rot-cylindr-coeff} (cf.\rf{spherical-WH} above) :
\be
ds^2 = -{\wti A} (dt)^2 - 2 {\wti E} dt\,d\vp + \frac{(d\eta)^2}{{\wti \D}} + 
{\wti D} (d\vp)^2 + \a^2 (r_{(+)} + |\eta|)^2 (dz)^2 \; ,
\lab{rot-WH}
\ee
where:
\br
{\wti A}(\eta) = A (r_{(+)} + |\eta|) \quad ,\quad
{\wti D}(\eta) = D (r_{(+)} + |\eta|) 
\nonu \\
{\wti E}(\eta) = E (r_{(+)} + |\eta|)  \quad ,\quad
{\wti \D}(\eta) = \D (r_{(+)} + |\eta|) \; , 
\lab{rot-WH-coeff}
\er
with $A,D,E,\D$ the same as in \rf{rot-cylindr-coeff}, and $r_{(+)}$ indicates the
outer horizon of \rf{rot-cylindr-metric}. From now on the bulk space-time indices
$\m,\n$ will refer to $(t,\eta,z,\vp)$ (instead of $(t,r,z,\vp)$).

The metric \rf{rot-WH}--\rf{rot-WH-coeff} represents two identical copies of the 
exterior region ($r > r_{(+)}$) of the cylindrically symmetric rotating black hole 
space-time with metric \rf{rot-cylindr-metric}, which are sewed together along 
the outer horizon $r=r_{(+)}$. The newly introduced coordinate $\eta$ 
($-\infty < \eta < \infty$)
will play the role of a planar radial-like coordinate normal w.r.t. the
\textsl{LL-brane} located on the horizon, which interpolates between two copies of
the exterior region of \rf{standard-spherical} (the two copies transform into
each other under the ``parity'' transformation $\eta \to - \eta$).

In the present case the \textsl{LL-brane} energy-momentum tensor
\rf{T-brane} has again the form \rf{T-S-0} with surface energy-momentum tensor
(cf. Eq.\rf{T-S-brane} above; now we have $D=p+2=4$) :
\be
S^{\m\n} = - \frac{\chi}{2a_0}\,\frac{|\g|}{|\g^2 -\om^2/\a^2|}
\llb - \pa_\t X^\m \pa_\t X^\n + \g^{ij} \pa_i X^\m \pa_j X^\n 
\rrb_{t=\t,\,\eta=0,\,z =\s^1,\,\vp=\s^2 + \t\,\om/\g} \; , 
\lab{T-S-brane-cylindr}
\ee
where \rf{cylindr-ansatz} and \rf{cylindr-horizon-dragging} are taken into
account. Here once again the dynamical \textsl{LL-brane} tension $\chi$
turns out to be {\em constant} unlike the exponential ``inflation/deflation''
\rf{chi-eq-cylindr} of the tension of test \textsl{LL-brane} moving in a fixed 
cylindrical black hole background. The proof is completely analogous to the
one given above for the spherically symmetric case.

As in the spherically symmetric case we separate in Einstein equations
\rf{Einstein-Maxwell-eqs} explicitly the terms contributing to $\delta$-function 
singularities on the \textsl{LL-brane} world-volume (obtaining Eqs.\rf{E-M-eqs}),
where again only the $(\m\n)=(\eta \eta)$ equation contains non-zero
$\delta$-function contributions, \textsl{i.e.}, arriving at Eq.\rf{E-M-eqs-eta}.
In the present case of cylindrically symmetric geometry \rf{rot-WH}--\rf{rot-WH-coeff}
Eq.\rf{E-M-eqs-eta} yields (taking into account the explicit form 
\rf{T-S-brane-cylindr} of the \textsl{LL-brane} surface energy-momentum tensor) :
\be
-\h \pa_\eta^2 \D (r_{(+)} + |\eta|) = 
8\pi\chi\,\frac{|\g|}{|\g^2 - \om^2/\a^2|}\,\d(\eta) 
+ \mathrm{non-singular ~terms} \, ,
\lab{E-M-eqs-eta-delta}
\ee
\be
\mathrm{i.e.} \quad
\pa_\eta \D (r_{(+)} + |\eta|)\bv_{\eta \to +0} 
- \pa_\eta \D (r_{(+)} + |\eta|)\bv_{\eta \to -0} = 
- 16\pi\chi\,\frac{|\g|}{|\g^2 - \om^2/\a^2|} \; ,
\lab{israel-junction-cylindr}
\ee
or, equivalently:
\be
\pa_r \D (r)\bv_{r=r_{(+)}} = - 8\pi\chi\,\frac{|\g|}{|\g^2 - \om^2/\a^2|}\, .
\lab{parameter-matching-cylindr}
\ee
Eq.\rf{parameter-matching-cylindr} provides a relation between the parameters 
$\a,\, b,\, c$ of the cylindrical rotating wormhole and the dynamical 
tension $\chi$ of the wormhole-generating \textsl{LL-brane}.
Here by construction the l.h.s. of \rf{parameter-matching-cylindr} is
strictly positive since $r_{(+)}$ is the outer horizon of the original metric
\rf{rot-WH}--\rf{rot-WH-coeff}, therefore, again as in the spherically
symmetric case the dynamical \textsl{LL-brane} tension $\chi$ at the
wormhole throat turns out to be {\em negative}.

\section{Traversability Considerations}

Let now briefly consider the dynamics of a test point particle moving in the
gravitational field of the cylindrically symmetric rotating wormhole
constructed in the previous Section. The analysis follows the lines of the 
standard procedure (see \textsl{e.g.} \ct{weinberg}).
The pertinent reparametrization invariant test-particle action reads:
\be
S_{{\rm particle}} = \h \int d\l \Bigl\lb \frac{1}{e}\xdot^\m \xdot^\n G_{\m\n}(x)
- e m^2\Bigr\rb  \; ,
\lab{particle-action}
\ee
where $G_{\m\n}(x) \equiv G_{\m\n}(t,\eta,z,\vp)$ is given by 
\rf{rot-WH}--\rf{rot-WH-coeff} and $e$ denotes the ``einbein''.
Variation w.r.t. $e$ yields the ``mass-shell'' constraint equation:
\be
-{\wti A} \tdot^2 - 2 {\wti E} \tdot \vpdot + {\wti D} \vpdot^2 + 
\a^2 (r_{(+)} + |\eta|)^2 \zdot^2 + \frac{1}{{\wti \D}}\,\etadot^2 + e^2 m^2 = 0 \; ,
\lab{mass-shell-eq}
\ee
with ${\wti A},\,{\wti D},\,{\wti E},\,{\wti \D}$ defined through
\rf{rot-WH-coeff} and \rf{rot-cylindr-coeff}.
We have also three Noether conserved quantities  -- energy $\cE$, axial angular 
momentum $\cJ$ and momentum $P_z$ along the $z$-axis:
\be
\cE = \frac{1}{e}\bigl( {\wti A}\tdot + {\wti E}\vpdot\Bigr) \;\; ,\;\;
\cJ = \frac{1}{e}\bigl( - {\wti E}\tdot + {\wti D}\vpdot\Bigr) \;\; ,\;\;
P_z = \frac{1}{e} \a^2 (r_{(+)} + |\eta|)^2 \zdot \; .
\lab{noether-eqs}
\ee
Solving \rf{noether-eqs} for $\tdot,\,\vpdot,\,\zdot$, substituting into 
\rf{mass-shell-eq} and employing the particle's proper-time $s$ instead of the 
generic evolution parameter$\l$ (the relation in the present case being given by
$ds/d\l=em$) we obtain the equation for the particle motion along
$\eta$ -- normal w.r.t. the wormhole throat:
\be
\eta^{\pr\, 2} + \cV_{\rm eff} (|\eta|) = \cE_{\rm eff} \; ,
\lab{particle-eq}
\ee
where:
\be
\cV_{\rm eff} (|\eta|) \equiv {\wti \D}(|\eta|)
\llb 1 + \frac{1}{m^2 (r_{(+)} + |\eta|)^2}\,\(\frac{P_z^2}{\a^2} +
\frac{\(\g\cJ - \cE \om/\a^2\)^2}{\(\g^2 - \om^2/\a^2\)^2}\)\rrb \; ,
\lab{V-eff}
\ee
\be
\cE_{\rm eff} \equiv \frac{\(\g\cE - \om\cJ\)^2}{m^2\(\g^2 - \om^2/\a^2\)^2} \; .
\lab{E-eff}
\ee
The ``effective potential'' $\cV_{\rm eff}$ \rf{V-eff} is strictly positive
for each $\eta \neq 0$ and growing with the following behavior
for small $\eta$ (\textsl{i.e.}, around the wormhole throat) and large $\eta$, 
respectively:
\be
\cV_{\rm eff} (|\eta|) 
\left\{ \begin{array}{ll} 
\simeq \mathrm{const}\, |\eta|  & \quad \mathrm{for}\;\; \eta \to 0 \\
\simeq \mathrm{const}\, \eta^2  & \quad \mathrm{for}\;\; \eta \to \pm \infty 
\end{array} \right. \; .
\lab{V-eff-asymptot}
\ee
Therefore, there is a whole range of values of the ``effective energy''
$\cE_{\rm eff}$ \rf{E-eff} for which the test particle periodically traverses 
the wormhole between the ``turning points'' $\pm \eta_{{\rm turning}}$ 
($\cV_{\rm eff} (|\pm \eta_{{\rm turning}}|) = \cE_{\rm eff}$) within a {\em
finite} amount of its proper time $s$.

On the other hand, for a static observer on either side of the wormhole
(which by construction is a copy of the exterior region of a black hole
beyond the outer horizon) the throat  looks the same as a black hole horizon, 
therefore, it would take an infinite amount of the ``laboratory'' time $t$ for a 
test particle to reach the wormhole throat. Thus, when we say that we
have constructed {\em traversable} wormholes via \textsl{LL-branes}, we have
in mind traversability w.r.t. {\em proper time} of travellers.

\section{Conclusions}

In the present paper we have provided a systematic general scheme to construct
self-consistent spherically symmetric or rotating cylindrical wormhole solutions via
\textsl{LL-branes}, such that the latter occupy the wormhole throats and
match together two copies of exterior regions of spherically symmetric or
rotating cylindrical black holes (the regions beyond the outer horizons).

As a particular case, the matching of two exterior regions of Schwarzschild
space-time at the horizon surface $r=2m$ through a \textsl{LL-brane} is
indeed the self-consistent realization of the original Einstein-Rosen ``bridge'',
namely, it requires the presence of a \textsl{LL-brane} at $r=2m$ -- 
a feature not recognized in the original Einstein-Rosen work \ct{einstein-rosen}
(see {\em Appendix}).

The main result here is the construction of self-consistent rotating
wormholes with rotating \textsl{LL-brane} as their sources sitting at the
wormhole throats. The surface tension of the \textsl{LL-brane} is an
additional brane degree of freedom, which assumes negative values on-shell
in all cases -- both for spherically symmetric as well as for rotating
cylindrical wormholes, but it can be of arbitrary small magnitude.

The class of spherically symmetric and rotating cylindrical wormhole solutions
produced by \textsl{LL-brane} constructed above combine the features of the 
original Einstein-Rosen ``bridge'' manifold \ct{einstein-rosen} (wormhole throat 
located at horizon) with the feature ``charge without charge'' of Misner-Wheeler 
wormholes \ct{misner-wheeler}. There exist several other types of physically
interesting wormhole solutions in the literature generated by different
types of matter and without horizons. For a recent discussion, see 
ref.\ct{lemos-bronnikov} and references therein.

The geodesic equations for test particles traversing the throats of the
presently constructed wormholes have been briefly studied. We have found
that it requires a {\em finite proper time} for a travelling observer to
pass from one side of the wormhole to the other, so that ``traversability''
for the presently constructed wormholes via \textsl{LL-branes} is understood
as traversability w.r.t. the proper time of travelling observers.

\section{Appendix: The Original Einstein-Rosen ``Bridge'' Needs a Lightlike Brane
For Consistency}

Here we briefly examine the first explicit wormhole construction proposed by 
Einstein and Rosen \ct{einstein-rosen} which is usually referred to as 
``Einstein-Rosen bridge''. As we will see in what follows, the Einstein-Rosen 
``bridge'' solution in terms of the original coordinates introduced in
\ct{einstein-rosen} (Eq.\rf{E-R-metric} below) {\em does not} satisfy the vacuum
Einstein equations due to an ill-defined $\d$-function contribution at the
throat appearing on the r.h.s. -- a would-be ``thin shell'' matter energy-momentum
tensor (see Eq.\rf{ricci-delta} below). The fully consistent formulation of the 
original Einstein-Rosen ``bridge'', namely, two identical copies of the exterior
Schwarzschild space-time region matched along the horizon must include a
gravity coupling to a \textsl{LL-brane}. In fact, the Einstein-Rosen ``bridge''
wormhole is a particular case  $(e=0)$ 
of our construction of Reissner-Nordstr{\"o}m wormhole via \textsl{LL-brane}
presented above in Section 4 (cf. Eqs.\rf{israel-junction}--\rf{M-Schw}). 
Here we will study separately the Einstein-Rosen ``bridge'' construction
because of its historic importance.

Let us start with the coordinate system proposed in \ct{einstein-rosen},
which is obtained from the original Schwarzschild coordinates by defining
$u^2=r-2m$, so that the Schwarzschild metric becomes:
\be
ds^2 = - \frac{u^2}{u^2 + 2m} (dt)^2 + 4 (u^2 + 2m)(du)^2 +
(u^2 + 2m)^2 \( (d\th)^2 + \sin^2 \th \,(d\vp)^2\) \; .
\lab{E-R-metric}
\ee
Then Einstein and Rosen ``double'' the exterior Schwarzschild space-time region 
($r>2m$) by letting the new coordinate $u$ to vary between $-\infty$ and $+\infty$
(\textsl{i.e.}, we have the same $r\geq 2m$ for $\pm u$). The two
Schwarzschild exterior space-time regions must be matched at the horizon $u=0$.

At this point let us note that the notion of ``Einstein-Rosen bridge'' in
\textsl{e.g.} ref.\ct{MTW}, which uses the Kruskal-Szekeres manifold,
is not equivalent to the original construction in \ct{einstein-rosen},
\textsl{i.e.}, two identical copies of the exterior Schwarzschild space-time region 
matched along the horizon. The two regions in
Kruskal-Szekeres space-time corresponding to the outer Schwarzschild space-time 
region ($r>2m$) and labeled $(I)$ and $(III)$ in \ct{MTW} are generally
{\em disconnected} and share only a two-sphere (the angular part) as a 
common border ($U=0, V=0$ in Kruskal-Szekeres coordinates), whereas in the 
original Einstein-Rosen ``bridge'' construction the boundary between the two 
identical copies of the outer four-dimensional Schwarzschild space-time region 
($r>2m$) is a three-dimensional hypersurface $(r=2m)$.

In our wormhole construction above (Section 4) we have used a different new
coordinate $\eta \in (-\infty,+\infty)$ to describe the two copies of the 
exterior (beyond the outer horizon) space-time regions. In the Schwarzschild case 
we have $|\eta| = r - 2m$ and, accordingly, the Schwarzschild metric
describing both copies becomes:
\be
ds^2 = - \frac{|\eta|}{|\eta| + 2m} (dt)^2 + \frac{|\eta| + 2m}{|\eta|}(d\eta)^2 +
(|\eta| + 2m)^2 \( (d\th)^2 + \sin^2 \th \,(d\vp)^2\)
\lab{our-schw-metric}
\ee
(one can use instead the Eddington-Finkelstein coordinate system; see below).
Due to the non-smooth dependence of the metric \rf{our-schw-metric} on $\eta$ 
via $|\eta|$ it is obvious that the terms in $R_{\m\n}$ containing second order 
derivative w.r.t. $\eta$ will generate $\d (\eta)$-terms on the l.h.s. of
the pertinent Einstein equations. It is precisely due to the gravity
coupling to \textsl{LL-brane}, that the corresponding \textsl{LL-brane}
surface stress-energy tensor on the r.h.s. of Einstein equations matches the
delta-function contributions on the l.h.s.. 
In particular, calculating the scalar curvature of the metric \rf{our-schw-metric}
we obtain the well-defined non-zero distributional result: 
\be
R = - \frac{1}{m} \d(\eta) \; .
\lab{R-delta}
\ee

The relation between both metrics \rf{E-R-metric} and \rf{our-schw-metric}
is a non-smooth coordinate transformation: 
\be
u =
\left\{\begin{array}{ll} 
\sqrt{\eta}  & \quad \mathrm{for}\;\; \eta \geq 0 \\
-\sqrt{-\eta} & \quad \mathrm{for}\;\; \eta \leq 0
\end{array} \right. \quad \mathrm{i.e.}\;\; u^2 = |\eta| \; .
\lab{u-eta}
\ee
Then the question arises as to whether one can see the presence of the 
\textsl{LL-brane} also in the Einstein-Rosen coordinates. The answer is ``yes'',
but with the important disclaimer that the Einstein-Rosen coordinate $u$ is 
{\em not appropriate} to describe the ``bridge'' at the throat $u=0$ as it
leads to an ill-defined $\d$-function singularity (Eq.\rf{ricci-delta} below).

To this end let us consider the Levi-Civita identity 
(see \textsl{e.g.} \ct{frankel}):
\be
R^0_0 = - \frac{1}{\sqrt{-g_{00}}} \nabla^2 \(\sqrt{-g_{00}}\)
\lab{levi-civita-id}
\ee
valid for any metric of the form 
$ds^2 = g_{00} (r) (dt)^2 + h_{ij}(r,\th,\vp) dx^i dx^j$ and where $\nabla^2$
is the three-dimensional Laplace-Beltrami operator
$\nabla^2 =\frac{1}{\sqrt{h}}\partder{}{x^i}\(\sqrt{h}\,h^{ij}\partder{}{x^j}\)$.
The Einstein-Rosen metric \rf{E-R-metric} solves $R^0_0 = 0$ for all $u\neq 0$.
However, since $\sqrt{-g_{00}} \sim |u|$ as $u \to 0$ and since
$\frac{\pa^2}{{\pa u}^2} |u| = 2 \d (u)$, Eq.\rf{levi-civita-id} tells us that:
\be
R^0_0 \sim \frac{1}{|u|} \d (u) \sim \d (u^2) \; ,
\lab{ricci-delta}
\ee
and similarly for the scalar curvature $R \sim \frac{1}{|u|} \d (u) \sim \d (u^2)$.
From \rf{ricci-delta} we conclude that: 

(i) The explicit presence of matter on the throat is missing in the original
formulation \ct{einstein-rosen} of Einstein-Rosen ``bridge''. 

(ii) The coordinate $u$ in \rf{E-R-metric} is {\em inadequate} 
for description of the original Einstein-Rosen ``bridge'' at the throat due to 
the {\em ill-definiteness} 
of the r.h.s. in \rf{ricci-delta}.

(iii) One should use instead the coordinate $\eta$ as in \rf{our-schw-metric}
(or as in \rf{our-EF-metric}--\rf{our-EF-metric-coeff} below), 
which provides the consistent construction of the original Einstein-Rosen 
``bridge'' manifold as a spherically symmetric wormhole with Schwarzschild geometry 
produced via lightlike brane sitting at its throat in a self-consistent formulation,
namely, solving Einstein equations with a surface stress-energy tensor
of the lightlike brane derived from a well-defined world-volume brane action.
Moreover, the mass parameter $m$ of the Einstein-Rosen ``bridge'' is not a
free parameter but rather is a function of the dynamical \textsl{LL-brane}
tension (Eq.\rf{M-Schw}).

Let us also describe the construction of Einstein-Rosen ``bridge'' wormhole
using the Finkelstein-Eddington coordinates for the Schwarzschild metric 
\ct{EFM} (see also \ct{MTW}):
\be
ds^2 = - A(r) (dv)^2 + 2 dv\,dr + r^2 \llb (d\th)^2 + \sin^2\th (d\vp)^2\rrb
\quad ;\quad A(r) = 1 - \frac{2m}{r} \; .
\lab{EF-metric}
\ee
The advantage of the metric \rf{EF-metric} over the metric in standard
Schwarzschild coordinates is that both \rf{EF-metric} as well as the
corresponding Christoffel coefficients {\em do not} exibit coordinate
singularities on the horizon $(r=2m)$.

Let us introduce the following modification of \rf{EF-metric}
(cf.\rf{our-schw-metric} above) :
\be
ds^2 = - {\wti A} (\eta) (dv)^2 + 2 dv\,d\eta + 
{\wti r}^2(\eta) \llb (d\th)^2 + \sin^2\th (d\vp)^2\rrb \; ,
\lab{our-EF-metric}
\ee
where:
\be
{\wti A} (\eta) = A (2m + |\eta|) = \frac{|\eta|}{|\eta| + 2m} \quad ,\quad
{\wti r}(\eta) = 2m + |\eta| \; .
\lab{our-EF-metric-coeff}
\ee
The metric describes two identical copies of Schwarzschild {\em exterior} region
($r > 2m$) in terms of the Eddington-Finkelstein coordinates, which correspond to 
$\eta >0$ and $\eta <0$, respectively, and which are ``glued'' together at the 
horizon $\eta = 0$ (\textsl{i.e.}, $r=2m$), where the latter will serve as a throat
of the overall wormhole solution. 

We will show that the metric \rf{our-EF-metric}--\rf{our-EF-metric-coeff} is a 
self-consistent solution of Einstein equations:
\be
R_{\m\n} - \h G_{\m\n} R = 8\pi T^{(brane)}_{\m\n}
\lab{Einstein-eqs}
\ee
derived from the action describing bulk ($D=4$) gravity coupled to a \textsl{LL-brane}:
\be
S = \int\!\! d^4 x\,\sqrt{-G}\,\frac{R(G)}{16\pi}\; +\; S_{\mathrm{LL}} \; ,
\lab{E-M-LL-4}
\ee
where $S_{\mathrm{LL}}$ is the \textsl{LL-brane} world-volume action \rf{LL-brane-chi}
with $p=2$.

Using as above the simplest non-trivial ansatz for the \textsl{LL-brane} embedding 
coordinates $X^\m \equiv (v,\eta,\th,\vp)=X^\m (\s)$: 
\be
v = \t \equiv \s^0 \;\; , \;\; \eta = \eta (\t) \;\; , \;\; \th^1 \equiv \th = \s^1
\;\;,\;\; \th^2 \equiv \vp = \s^2 \; ,
\lab{X-embed-0}
\ee
the pertinent \textsl{LL-brane} equations of motion yield (in complete
analogy with \rf{r-const}--\rf{horizon-standard}) :
\be
\eta (\t) = 0 \quad ,\quad \pa_\t \chi + \chi \Bigl\lb \h \pa_\eta {\wti A} 
+ 2a_0 \,\pa_\eta \ln {\wti r}^2 \Bigr\rb_{\eta=0} = 0 \, .
\lab{eta-chi-eqs}
\ee
As above, the first Eq.\rf{eta-chi-eqs} (horizon ``straddling'' by the
\textsl{LL-brane}) is obtained from the constraint 
equations \rf{gamma-eqs-0}, whereas the second Eq.\rf{eta-chi-eqs} results from 
the geodesic \textsl{LL-brane} equation for $X^0\equiv v$ \rf{X-eqs-0} due to 
the embedding \rf{X-embed-0}. Here again as in Section 4, the problem with
the discontinuity in the Christoffel coefficients accross the horizon ($\eta=0$)
is resolved following the approach in ref.\ct{Israel-66} (see also the regularization
approach in ref.\ct{BGG}, Appendix A), \textsl{i.e.}, we need to
take the mean value w.r.t. $\eta=0$ yielding zero (since both ${\wti A}$ and 
${\wti r}$ depend on $|\eta|$). Therefore, once again as in Section 4 we
find that the dynamical \textsl{LL-brane} tension $\chi$ turns into an integration
constant on-shell.

Taking into account \rf{X-embed-0}--\rf{eta-chi-eqs} the \textsl{LL-brane}
energy momentum tensor \rf{T-brane} derived from the world-volume action 
\rf{LL-brane-chi} becomes (cf. Eq.\rf{T-S-brane}):
\be
T_{(brane)}^{\m\n} = S^{\m\n}\,\d (\eta) \quad ,\quad
S^{\m\n} = \frac{\chi}{2a_0}\llb\pa_\t X^\m \pa_\t X^\n -
2a_0 G^{ij} \pa_i X^\m \pa_j X^\n \rrb_{v=\t,\,\eta=0,\,\th^i =\s^i} 
\lab{T-S-brane-0}
\ee
where $G^{ij}$ is the inverse metric in the $(\th,\vp)$ subspace and $a_0$
indicates the integration constant parameter arising in the
\textsl{LL-brane} world-volume dynamics (Eq.\rf{gamma-eqs-u}).

Now we turn to the Einstein equations \rf{Einstein-eqs} where again as in
Section 4 (cf. Eqs\rf{E-M-eqs}) we explicitly separate the terms contributing to 
$\d$-function singularities on the l.h.s.:
\br
R_{\m\n} \equiv \pa_\eta \G^{\eta}_{\m\n} - \pa_\m \pa_\n \ln \sqrt{-G}
+ \mathrm{non-singular ~terms}
= 8\pi \Bigl( S_{\m\n} - \h G_{\m\n} S^{\l}_{\l}\Bigr) \d (\eta) 
\lab{E-eqs-sing}
\er
Using the explicit expressions: 
\be
\G^{\eta}_{vv} = \h {\wti A}\pa_\eta {\wti A} \;\;,\;\;
\G^{\eta}_{v\,\eta} = -\h \pa_\eta {\wti A} \;\; ,\;\; \G^{\eta}_{\eta\eta} = 0 
\;\; , \;\; \sqrt{-G} = {\wti r}^2
\lab{Christoffel-EF}
\ee
with ${\wti A}$ and ${\wti r}$ as in \rf{our-EF-metric-coeff}, it is straighforward 
to check that non-zero $\d$-function contributions in $R_{\m\n}$ appear for 
$(\m\n)=(v\,\eta)$ and $(\m\n)=(\eta\eta)$ only. Using also the expressions
$S_{\eta\eta}=\frac{1}{2a_0}\chi$ and $S^\l_\l = - 2\chi$ 
(cf. Eq.\rf{T-S-brane-trace}) the Einstein Eqs.\rf{E-eqs-sing} yield for
$(\m\n)=(v\,\eta)$ and $(\m\n)=(\eta\eta)$ the following matchings of the
coefficients in front of the $\d$-functions, respectively:
\be
m = \frac{1}{16\pi|\chi|} \quad ,\quad m = \frac{a_0}{2\pi|\chi|} \; .
\lab{M-Schw-EF}
\ee
where as above the \textsl{LL-brane} dynamical tension must be negative.
Consistency between the two relations \rf{M-Schw-EF} fixes the value
$a_0 = 1/8$ for the integration constant $a_0$.

Thus, we recover the same expression for the Schwarzschild mass $m$
of the Einstein-Rosen ``bridge'' wormhole (as function of the dynamical 
\textsl{LL-brane} tension) in the Eddington-Finkelstein coordinates 
(first Eq.\rf{M-Schw-EF}) as in the standard Schwarzschild 
coordinates (Eq.\rf{M-Schw}). Moreover, unlike the previous treatment with
the standard Schwarzschild coordinates, there are no coordinate
singularities in the Christoffel coefficients \rf{Christoffel-EF}, so when
employing Eddington-Finkelstein coordinates there is no need to use mixed indices
(one covariant and one contravariant) in the Einstein equations unlike 
\rf{E-M-eqs-eta}.

In conclusion let us note that for the scalar curvature of the Einstein-Rosen
``bridge'' wormhole metric in Eddington-Finkelstein coordinates 
\rf{our-EF-metric}--\rf{our-EF-metric-coeff}
we obtain the same well-defined non-zero distributional result
as in the case with ordinary Schwarzschild coordinates \rf{R-delta}:
\be
R = - \frac{1}{m} \d(\eta) \; .
\lab{R-delta-EF}
\ee

\section*{Acknowledgments}
We are very grateful to Werner Israel and Eric Poisson for useful and
illuminating correspondence.
E.N. and S.P. are supported by Bulgarian NSF grant \textsl{DO 02-257}.
Also, all of us acknowledge support of our collaboration through the exchange
agreement between the Ben-Gurion University of the Negev (Beer-Sheva, Israel) and
the Bulgarian Academy of Sciences.


\end{document}